\documentclass[epj, final]{svjour}

\usepackage[english]{babel}
\usepackage{graphicx}
\usepackage[usenames]{color}
\usepackage{mathrsfs}
\usepackage{amssymb}
\usepackage{url}
\begin{document}

\newcommand{\etal}{\textit{et~al.}\
}   
\newcommand{\eg}{\textit{e.g.}\
}   
\newcommand{\ie}{\textit{i.e.}\
}   
\newcommand{\via}{via }
\let\oldarccos\arccos
\renewcommand{\arccos}[1]{\oldarccos\left(#1\right)}
\let\oldcos\cos
\renewcommand{\cos}[1]{\oldcos\left(#1\right)}
\newcommand{\cossq}[1]{\oldcos^2\left(#1\right)}
\let\oldsin\sin
\renewcommand{\sin}[1]{\oldsin\left(#1\right)}
\newcommand{\sinsq}[1]{\oldsin^2\left(#1\right)}
\newcommand{\ket}[1]{|#1\rangle}
\newcommand{\bra}[1]{\langle#1|}
\newcommand{\cc}[1]{c_{#1}^\dagger}
\newcommand{\cd}[1]{c_{#1}^{\phantom{\dagger}}}

\title{Coulomb blockade without potential barriers} 

\author{Gabriel Vasseur \and Dietmar Weinmann\thanks{\emph{e-mail:}
Dietmar.Weinmann@ipcms.u-strasbg.fr} \and Rodolfo A.\ Jalabert}

\institute{Institut de Physique et Chimie des Mat\'eriaux de Strasbourg,\thanks{UMR
7504 ULP-CNRS} 23 rue du L{\oe}ss, BP 43, 67034 Strasbourg cedex 2, France}

\date{\today}

\abstract{
We study transport through a strongly correlated quantum dot and show
that Coulomb blockade can appear even in the presence of perfect contacts. This
conclusion arises from numerical calculations of the conductance for a
microscopic model of spinless fermions in an
interacting chain connected to each lead \via a completely open channel. The
dependence of the conductance on the gate voltage shows well
defined Coulomb blockade peaks which are sharpened as the interaction strength
is increased. Our numerics is based on the embedding method
and the DMRG algorithm. We explain the emergence of Coulomb blockade
with perfect contacts by a reduction of the effective coupling matrix elements
between many-body states corresponding to successive particle numbers in the
interacting region. A perturbative approach, valid in the strong interaction
limit, yields an analytic expression for the interaction-induced suppression of
the conductance in the Coulomb blockade regime.
} 

\maketitle

\section{Introduction}

The discreteness of the electron charge blocks the transport through a small and
relatively isolated conductor (usually referred as a quantum dot) at sufficiently low temperature
and bias voltage. This peculiar phenomenon, known as Coulomb blockade, is a
paradigm in the physics of small condensed matter systems. The Coulomb blockade
can be lifted by tuning a gate voltage to a point of degeneracy, where the
energy cost for adding or removing an electron to the system vanishes. The
obvious signature of large conductance oscillations as a function of gate
voltage makes this phenomenon promising for applications, and at the same time a
privileged set-up to investigate interactions in confined systems
\cite{kouwenhoven1997,beenakker1991}.

A question that has been present since the beginning of the studies on Coulomb
blockade is: what are the minimum ingredients for it to be observed? The
repulsive inter-particle interaction is clearly one of them, as well as the
condition that the charging energy of the dot be greater than the thermal energy
$k_\mathrm{B} T$ or the applied bias $eV$. Having an almost isolated dot, connected to
conducting leads by weakly transmitting contacts like tunnel-barriers, has been
thought to be another essential ingredient. Various studies have been undertaken
to follow the evolution of the Coulomb blockade oscillations as the dot becomes
better coupled to the leads.

It was in the quantum Hall regime that Coulomb blockade effects in open (well
coupled) dots were initially investigated \cite{alphenaar1992,marmorkos1992}.
The existence of spatially separated edge channels made it possible to observe
conductance oscillations even in the case in which the point contacts at the
entrance of the dot had a conductance $G_\mathrm{B}$ larger than the quantum conductance
$G_0=e^2/h$.

In the absence of the quantum Hall effect, the regime of weak Coulomb blockade,
which is intermediate between that of strong Coulomb blockade ($G_\mathrm{B}\ll G_0$) and
the one of a large number of fully open channels ($G_\mathrm{B}\gg G_0$, where no Coulomb
blockade can be observed), has been studied \cite{aleiner2002,brouwer2005}. In
the regime of weak Coulomb blockade ($G_\mathrm{B}\sim G_0$), the precise behaviour of
the system depends on the way in which the value of the entrance conductance is
obtained. In the case of metallic islands coupled to leads \via tunnel-barriers,
$G_\mathrm{B}$ can be of the order of $G_0$ due to a large number of weakly transmitting
channels; while in semiconducting dots coupled to leads \via point contacts,
conductances $G_\mathrm{B}$ of the order of $G_0$ are obtained by increasing the
transmission of only a few channels.

Nazarov has shown that charge quantization may persist when the dot is connected
to the leads \via arbitrary conductors \cite{nazarov1999}, as for example
disordered metallic wires, instead of tunnel-barriers. In the general case of a
number of not perfectly transmitting channels, the effective charging energy is
exponentially small (scaling as $\exp{\left(-G_\mathrm{B}/G_0\right)}$).
However, the charging energy vanishes if one of the channels is perfectly
transmitting.

It is important to make the distinction between perfect transmission and perfect
contact. The notion of transmission of the barrier, which should be ``measured''
between two electrodes, may not be a relevant concept when the barrier is between a dot and
an electrode. In this case a perfect contact does not guarantee the absence of
reflection, since electrons can be coherently backscattered inside the dot.

Recent works \cite{golubev2004,brouwer2005} considering open dots have taken
into account the coherent backscattering induced by the dot itself. They found
that interactions do not modify the average conductance when all the channels
are either open or closed (case of ideal contacts). This conclusion is in
agreement with an experiment performed on rather large chaotic dots
\cite{huibers1998}.

Experimentally, Coulomb blockade physics has been observed recently in very
small silicon based MOSFETs where the tunnel-barriers are not built-in
\cite{boehm2005,max}. It is likely that the Coulomb blockade features of these
samples are due either to diffusive leads or to electrostatic potential barriers
created between the dot and the leads. It is, however, of interest to study
whether it is really necessary to have poor contacts for obtaining the Coulomb
blockade. For dots of reduced dimensionality and containing a low number of
electrons, we expect electronic correlations to be of crucial importance.
Therefore we may ask whether Coulomb blockade can arise even with perfect
contacts, due only to strong correlations.

In order to address this question, it is important to go beyond the usual
treatment of the interactions in the dot based on the capacitive charging
energy. Microscopic models of correlated chains attached to semi-infinite leads
have recently been investigated \cite{nisikawa_cond-mat,bohr2006,schneiderCM}.
For good but not perfect contacts, it has been found that Coulomb blockade-like
features are reinforced by the interactions.

In this paper, we present a study of a microscopic model for a strongly
correlated quantum dot and show that Coulomb blockade arises even in the
presence of perfect contacts. 

The model and the method are introduced in section \ref{methodandmodel}. The
numerical results presented in section \ref{numerics} show that Coulomb blockade
appears as the interaction strength increases, even in the absence of potential
barriers between the dot and the leads. The features of the peaks and valleys
are analyzed in detail in section \ref{analytics} using perturbative approaches
at weak transmission. In the concluding section, we discuss the implications of
our findings and possible extensions of our model in order to approach the
current experimental setups.

\section{Embedding method for the conductance of a one-dimensional model
\label{methodandmodel}}

Most of the analytical approaches to Coulomb blockade rely on the so-called
constant charging model, where correlations are completely disregarded. The only
electron-electron interaction term considered is the capacitive charging energy
of the quantum dot. While various refinements, such as the Random Phase
Approximation \cite{blanter1997}, have been developed to describe the case of
small dots with poor screening, treating electronic correlations to all orders
is obviously out of reach in any analytical approach. Numerically, various
schemes for obtaining the addition spectrum of a quantum dot have been used
\cite{cohen1999}. However, these calculations of ground state energies cannot
capture the richness and complexity of a transport problem, especially for open
quantum dots.

Even numerically, the transport through strongly correlated systems has remained
an intractable problem until recently. The development of the so-called
embedding method \cite{molina2003,molina2004a} has been an important advance in
this difficult problem. Relating the transport properties of a quantum dot with
the thermodynamic properties of the combined system (sample $+$ leads, as we
explain in the sequel), allows to extract the conductance through the dot from
the numerically accessible ground state energy of the combined system.

The embedding method can readily be used for spinless fermions (spin-polarized
electrons) in one-dimensional chains. While the method has been successfully
generalized to spin-half electrons \cite{future}, the problem of spinless
fermions remains more easily tractable. For these reasons we restrict ourselves
in this work to one-dimensional spinless fermions. It is expected that the
qualitative aspects of our results apply also in higher dimensions and once we
take the electron spin into account.

\begin{figure}[tbp]
\resizebox{1.00\columnwidth}{!}{%
\includegraphics{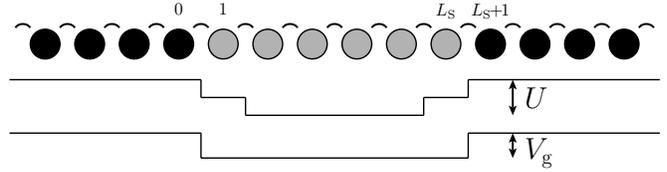}
}
\caption{The central region of an infinitely long, one-dimensional chain. The
particles experience nearest neighbour interaction of strength $U$ on sites $1$
to $L_\mathrm{S}$. Below the chain are sketched the compensating potential of
the interaction (see text) and the potential due to the gate voltage
$V_\mathrm{g}$.
} 
\label{system} 
\end{figure}

We consider a model of spinless fermions in a chain, as
sketched in Fig.\ \ref{system}. The dot corresponds to a region of length
$L_\mathrm{S}$ in which the particles are interacting. In the rest of the chain,
representing the leads of the standard experimental set-up, the particles do not
interact. Therefore, far away from the dot, a Fermi energy $E_\mathrm{F}$ is well-defined.
We fix the latter to the centre of the band, corresponding to half filling.
The Hamiltonian of the whole system reads 
\begin{equation}\label{eq:hamiltonian}
H = H_\mathrm{K} + H_U + H_\mathrm{G} ,
\end{equation}
where
\begin{equation}
H_\mathrm{K} = -\sum_{i=-\infty}^{\infty} ( \cc{i} \cd{i+1} + h.c. )
\end{equation}
stands for the kinetic energy of an ideal one-dimensional chain.
Here, $c_i$ annihilates a particle on site $i$. Setting the hopping amplitudes
in $H_\mathrm{K}$ to unity defines our energy scale. 

The dot is represented by the sites $i=1$ to $L_\mathrm{S}$ (the lattice constant is set
to unity). 
The inter-particle interaction
\begin{equation}
H_U = U \sum_{i=1}^{L_\mathrm{S}-1} (\hat{n}_i - 1/2) (\hat{n}_{i+1} -1/2)
\end{equation}
acts only in this region and is assumed to be restricted to nearest
neighbours (with strength $U$ and $\hat{n}_i = \cc{i} \cd{i}$). 
$H_U$ includes a one-body compensating potential (due to the presence of the
$1/2$'s, see Fig.\ \ref{system}) which ensures, in the absence of gate voltage,
particle-hole symmetry even at $U\neq 0$. It can be seen as a positive
background that prevents the particles from leaking out of the dot region, in
spite of the interaction. Additional particles can then be attracted in the dot
by the one-body potential
\begin{equation}
H_\mathrm{G} = -V_\mathrm{g} \sum_{i=1}^{L_\mathrm{S}} \hat{n}_i
\end{equation}
describing the effect of an applied gate voltage $V_\mathrm{g}$. 

The relative isolation of a dot is usually achieved by connecting it to the
leads \via weak links or tunnel-barriers (\ie by choosing a smaller hopping
amplitude or a high on-site potential), but we do not take any of these
approaches in this work. In contrast, in our case, {\it the contacts between the dot
and the leads are perfect}. The dot is therefore only 
defined by the region where the  electron-electron interactions and the gate
voltage are present.

We use the embedding method which allows to determine the modulus $|t|$ of the
effective transmission amplitude $t=|t|e^{i\alpha}$ of the system at the Fermi
energy of the electrodes, taking fully into account the effects of electronic
correlations. This thereby yields the zero-temperature, linear-response
conductance $g=|t|^2$ (in units of $e^2/h$). The modulus $|t|$ is obtained from
the persistent current flowing in a ring formed by the interacting dot together
with an infinitely long, non-interacting lead.

The relation between the persistent current and the conductance is based on two
facts, which are both realized in the limit of an infinitely long lead. First,
the persistent current in a ring including a {\it one-body}
(non-interacting) scatterer only depends, in this limit, on the modulus $|t|$
of the transmission amplitude of the scatterer. Second, it is only in
the limit of an infinitely long lead that a ring containing an {\it interacting
part} behaves as a Fermi liquid, and that a one-body transmission amplitude $t$
can be attributed to the interacting region attached to leads \cite{molina2005}.

The study of closed, finite-size systems thus allows to obtain the Landauer
conductance of the dot in the setup described by the Hamiltonian
(\ref{eq:hamiltonian}), provided that an extrapolation to infinite size is
performed. We are able to perform reliable extrapolations for an interacting dot
of $L_\mathrm{S}=6$, even near the resonances, where the extrapolation procedure
is most difficult \cite{molina2004a}. Therefore, besides the results for the
physics of Coulomb blockade, this work also constitutes a test of the embedding
method in a computationally demanding situation.

To be specific, we use the phase sensitivity $\mathcal{D}= (L/2) |E_\mathrm{P} -
E_\mathrm{A}|$,
where $E_\mathrm{P,A}$ denotes the ground state energies of the ring for
periodic (antiperiodic) boundary conditions, and $L$ is the total size of the
ring. The extrapolation to $L \to \infty$ is carried out from data calculated for
different ring sizes. The limiting value $\mathcal{D}_{\infty}$ leads to the
effective transmission amplitude of the system at the Fermi energy
\cite{molina2003,rejec2003}
\begin{equation}
|t| = \sin{\mathcal{D}_{\infty}}.
\end{equation}
Similar approaches using the persistent current have also been recently proposed
\cite{favand1998,sushkov2001,meden2003}.

We take advantage of the efficiency of the Density Matrix Renormalization Group
algorithm (DMRG) \cite{DMRG_book,schmitteckert_thesis} which allows to calculate
$\mathcal{D}$ with very high precision, taking fully into account the electronic
correlations. 
The extrapolation can then usually be performed from the many-body ground state
energies of rings of up to 100 sites. Near the resonances, however, the
extrapolation can be quite difficult and we need to treat rings containing up to
300 sites, keeping up to 1000 states in the DMRG iterations.

\section{Interaction-induced conductance oscillations as a function of gate voltage
\label{numerics}}

\begin{figure}[tbp]
\resizebox{1.00\columnwidth}{!}{%
\includegraphics{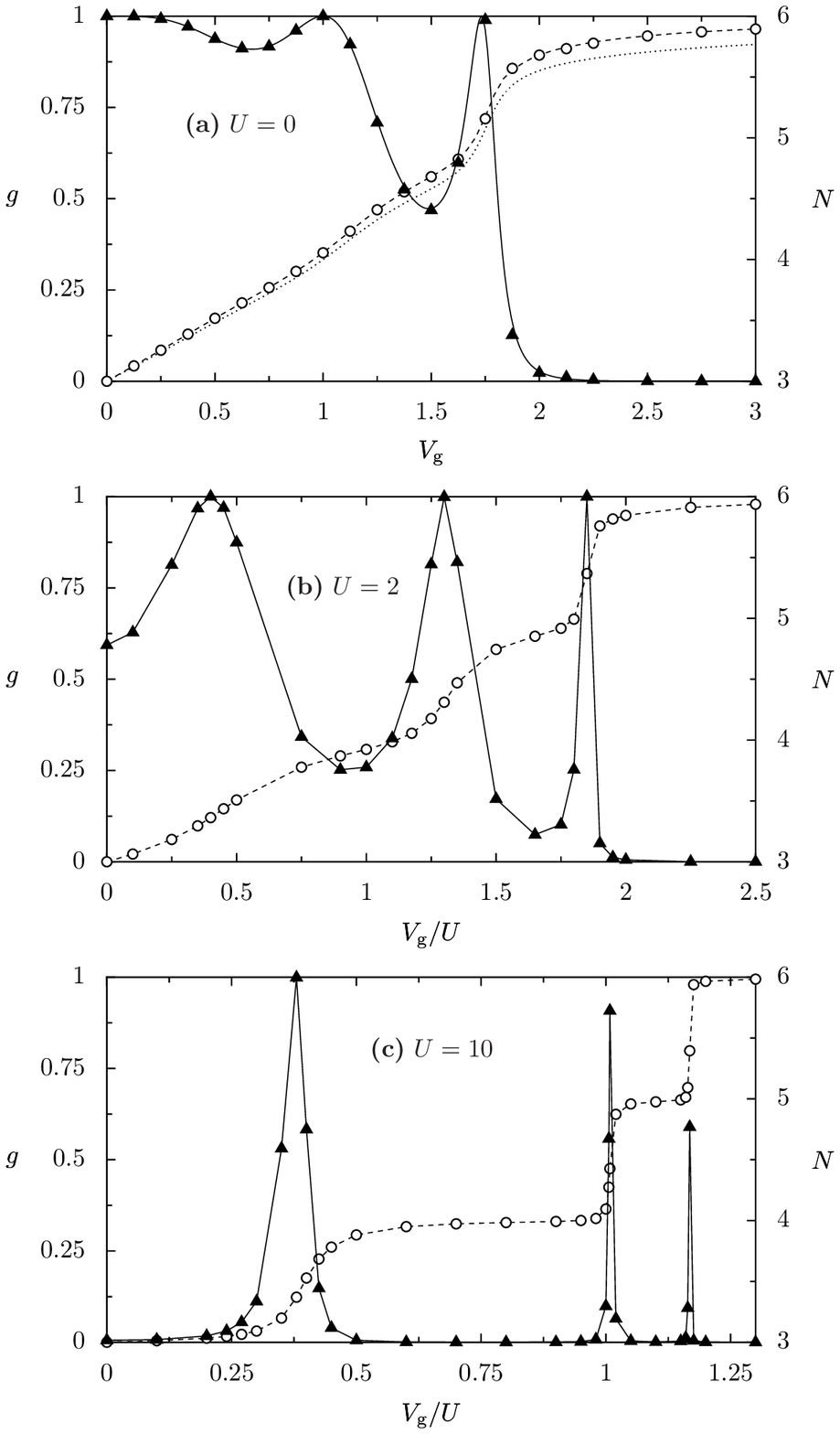}
}
\caption{Dimensionless conductance $g$ (triangles) and mean number of particles $N$
(circles) inside the dot (of $L_\mathrm{S}=6$ sites), as a function of the gate voltage
$V_\mathrm{g}$ without interaction {\bf(a)} and $V_\mathrm{g}/U$ for $U=2$
{\bf(b)} and $U=10$ {\bf(c)}, obtained by DMRG calculations. In the
non-interacting case {\bf(a)}, $g$ is known analytically (Eq.\ (\ref{ga}), solid
line). The Friedel sum rule (dotted line) gives $N$ approximately (see text)
while one-body numerics (dashed line) is in perfect agreement with the DMRG
points. In {\bf(b)} and {\bf(c)} lines are guides to the eye.}
\label{all} 
\end{figure}

The dimensionless conductance $g$ as a function of the gate voltage
$V_\mathrm{g}$ is shown in Fig.\ \ref{all}, together with the mean charge inside
the dot region. The three graphs shown are for a dot of length $L_\mathrm{S}=6$,
and different interaction strengths.

The non-interacting case ($U=0$, Fig.\ \ref{all}a) corresponds to free particles
in a chain with a potential well of length $L_\mathrm{S}$ and depth $V_\mathrm{g}$. The
structures in the conductance for this case can be understood from one-body
quantum transport through the well potential, and thus
$t(E_\mathrm{F})=|t|e^{i\alpha}$ can be readily calculated. The conductance is then
obtained as $g=|t|^2$ which, when $|V_\mathrm{g}| \leq 2$ and for
$E_\mathrm{F}=0$, can be written as
\begin{equation}
|t|^2 = \left ( \cossq{ f(V_\mathrm{g}) L_\mathrm{S}} + \frac{\sinsq{f(V_\mathrm{g})
L_\mathrm{S}}}{\sinsq{f(V_\mathrm{g})}} \right )^{-1} ,
\label{ga}
\end{equation}
with $f(V_\mathrm{g}) = \arccos{-V_\mathrm{g}/2}$. The result, plotted in Fig.\
\ref{all}a, is in perfect agreement with the numerical data obtained from the
embedding method. This is a stringent test of our method since the
non-interacting case does not represent a special situation for the DMRG
algorithm. Furthermore, the transmission phase $\alpha$ can be related to the
number $N$ of particles inside the dot by the Friedel sum rule
\cite{friedel1,friedel2}
\begin{equation}
N_\mathrm{d} = \alpha / \pi ,
\label{friedel}
\end{equation}
where $N_\mathrm{d}$ is the number of particles displaced by the dot.
$N_\mathrm{d}$ gives the
number $N$ of particles {\it inside} the dot only for a system with homogeneous
density, such that deviations between the analytic result (dotted line in Fig.\
\ref{all}a) and the DMRG data appear with increasing gate voltage, when part of
the displaced particles are in the leads. We also calculated $N$ from a direct
diagonalization of the one-body Hamiltonian, and find perfect agreement with the
DMRG data (see Fig.\ \ref{all}a, dashed line).

As one increases the interaction, {\it a well-defined peak structure appears}.
The electron-electron interaction suppresses the conductance between the peaks
which become sharper with increasing interaction strength. This tendency can be
observed already at moderate interaction strength $U=2$ (Fig.\ \ref{all}b), and
becomes very pronounced at strong interaction $U=10$, as can be seen in Fig.\
\ref{all}c.
In a one-body problem with symmetric contacts the conductance at resonance
reaches unity. In our strongly correlated problem the limited amount of data
points and the uncertainty of the extrapolation do not allow to settle the
precise values at resonance in Fig.\ \ref{all}c.

The emergence of a well-defined peak structure in the conductance $g$ is
accompanied by the appearance of a staircase dependence on $V_\mathrm{g}$ of the mean
particle number $N$ inside the dot region. {\it The plateaus appear at integer values
of the mean particle number}, and the steps in $N$ occur precisely at the gate
voltage values of the conductance peaks.    

Although the dot region is perfectly connected to the leads, these structures
are clearly due to the Coulomb blockade effect, and their features can be
identified with the properties of the usual Coulomb blockade conductance
oscillations. It therefore appears that the dot becomes effectively decoupled
from the leads by the electron-electron interaction. The underlying mechanism
will be discussed in the following sections. In particular, we will show in
section \ref{peaks} that when the interaction is strong enough, the peak
positions can be predicted precisely from the knowledge of the ground state
energies of the isolated dot system for different mean particle numbers. Similar
conclusions have been drawn from numerical studies of a chain containing 7
sites, but connected to the leads by reduced hopping matrix elements
\cite{bohr2006,schneiderCM}. In this case, a DMRG evaluation of the Kubo formula
and a DMRG study of the real-time dynamics have been performed showing a
reinforcement of Coulomb blockade features when the interaction strength is
increased. Including the spin degree of freedom leads to the appearance of
conductance plateaus in the Kondo regime. A numerical renormalization group
(NRG) study \cite{nisikawa_cond-mat} of a Hubbard chain of 4 sites with moderate
interaction strength and reduced coupling to the leads shows that the
corresponding features are smoothed when the coupling to the leads is increased.  

Since the full Hamiltonian is symmetric under the exchange of particles and
holes with a simultaneous reversal of the sign of the gate voltage, the results
for negative gate voltage can be obtained directly from the data presented in
Fig.\ \ref{all}. The conductance is symmetric with respect to the gate voltage,
$g(-V_\mathrm{g})=g(V_\mathrm{g})$, while the number of particles exhibits the property
$N(-V_\mathrm{g})=L_\mathrm{S}-N(V_\mathrm{g})$. Thus, as expected for a dot with $L_\mathrm{S}=6$
sites, there is a total of 6 peaks with the particle number increasing in steps
from 0 to 6. Systems with larger $L_\mathrm{S}$ are therefore expected to give
qualitatively similar results with $L_\mathrm{S}$ conductance peaks and
$L_\mathrm{S}$ charge steps. In the case of odd $L_\mathrm{S}$, the number of
conductance peaks is odd. Because the conductance is an even function of the
gate voltage, one of the conductance peaks must be at $V_\mathrm{g}=0$, in contrast to
the case of even $L_\mathrm{S}$, where $V_\mathrm{g}=0$ is always in the centre of a
valley between two conductance peaks. This observation allows to understand the
even-odd oscillations in the conductance of an interacting chain without gate
voltage \cite{molina2003} discussed in detail in Ref.\ \cite{molina2004}. The
oscillations of the conductance with the length of the interacting region are a
very general feature of interacting chains. A perturbative treatment of the
interactions \cite{oguri1999} indicates that they appear in Hubbard chains as
well. They may be related to experimentally observed conductance oscillations
with the length of mono-atomic chains \cite{smit2003}.

\section{Perturbative approaches for weak transmission
\label{analytics}
}

In order to develop our understanding of the physical features found
numerically,  we present an analytical study of the model for the regime of
strong interaction. The behaviour observed in Fig.\ \ref{all}, and discussed in
the previous section, indicates that the dot becomes decoupled from the leads
when the interaction strength is increased. We will show that this is indeed the
case. To this end, we treat the conductance through the system with a Fermi
golden rule approach, valid when the effective coupling between the dot and the
leads is small. In such a framework, the starting point is a dot region which is
isolated from the leads. The coupling between the dot and the leads is then
treated as a perturbation. In this case, the relevant coupling is not given by
the one-body hopping amplitude between the sites $0 \leftrightarrow 1$ and
$L_\mathrm{S} \leftrightarrow L_\mathrm{S}+1$, which in our model is always
unity, but by the matrix elements of the terms of the Hamiltonian
(\ref{eq:hamiltonian}) which couple the interacting dot region to the
semi-infinite leads, sandwiched between the full many-body states of the chain.
This coupling part of the Hamiltonian reads
\begin{equation}
H_\mathrm{C} = - ( \cc{0} \cd{1} + h.c. ) - ( \cc{L_\mathrm{S}} \cd{L_\mathrm{S}+1}
+ h.c.)
\end{equation}
and corresponds to $H_\mathrm{K}$ restricted to the two links between the dot and the
leads. We now evaluate the lowest order contribution in $H_\mathrm{C}$ to the conductance
of our system. In the case where the involved matrix elements of $H_\mathrm{C}$ are small
(compared to $U$), the conductance is dominated by this term.

The conductance is related to the transmission of electrons through the system.
Let us start by considering Fermi's golden rule for the transition rate 
\begin{equation}
{\gamma}_{i \to f} = \frac{2\pi}{\hbar} {| M_{f,i} |}^2  \delta(
E_f - E_i )
\label{FGR}
\end{equation}
of a particle through the dot region. The initial state
\begin{equation}
|i\rangle = | k^\mathrm{left} \rangle \otimes | 0(N) \rangle 
\label{i}
\end{equation}
describes a product state built from eigenstates of the isolated dot and the
semi-infinite leads. Here, we take the dot in its $N$-particle ground state $|
0(N) \rangle $ and an additional particle $| k^\mathrm{left} \rangle$ with
wavenumber $k$ in the left lead. In the final state
\begin{equation}
|f\rangle = | 0(N) \rangle \otimes | k^\mathrm{right} \rangle 
\label{f}
\end{equation}
the additional particle appears in the right lead while the dot returns to its
$N$-particle ground state. The effective transition matrix element is of second
order in $H_\mathrm{C}$, and can be written as 
\begin{equation}
M_{f,i} = \sum_{\alpha} \frac{\langle i | H_\mathrm{C} | \alpha \rangle \langle
  \alpha | H_\mathrm{C} | f \rangle}{E_i - E_\mathrm{\alpha}}.
\label{Mfi}
\end{equation}
Here, the sum runs over all intermediate states $|\alpha \rangle$ corresponding
to $N+1$ particles in the dot region and the Fermi vacuum in the leads or to $N-1$
particles in the dot and both leads occupied by an
extra particle. Summing the transition rate (\ref{FGR}) over the states of the
non-interacting leads in a small energy interval corresponding to an
infinitesimal bias voltage $V$, one can obtain the corresponding current and
thus the dimensionless conductance
\begin{equation}
g = 4 \pi^2 \rho^2 | M_{f,i} |^2 ,
\label{g}
\end{equation}
where $\rho$ is the density of states in the leads.

The approach is justified when the effective coupling given by the matrix
element in Eq.\ (\ref{Mfi}) is small. This is the case at large $U$ when the
energy denominator is large, even though we have perfect contacts and
$H_\mathrm{C}$ is not small. In addition, the matrix elements of $H_\mathrm{C}$
can be reduced due to the interaction-induced modification of the dot
wave-functions.

\subsection{Conductance in the limit of strong interactions\label{strongU}}

In the strong interaction limit ($U \gg 1, V_\mathrm{g}$), the ground state of the
isolated dot region for an even dot length $L_\mathrm{S}$ at half filling ($N=L_\mathrm{S}/2$) is
given by a charge density wave (or Mott insulator, where the particles occupy
alternating sites). In this case we can evaluate $M_{f,i}$, and hence obtain the
conductance (\ref{g}) in the corresponding conductance valley. 

In this section, we treat not only the coupling between the dot and the leads,
but also all of the hopping terms $H_\mathrm{K}$ as a perturbation to the other terms
$H_0 = H_U + H_\mathrm{G}$ of the Hamiltonian. This corresponds to an expansion in
$1/U$ and follows the spirit of the large $U$ expansions of the persistent
current presented in Ref. \cite{selva2000}. In the absence of $H_\mathrm{K}$, the two
realizations of the Mott insulator with the particles on the even sites
$|\psi_0^{\mathrm{e}}\rangle = \cc{2} \cc{4} \ldots \cc{L_\mathrm{S}} | 0
\rangle$ and on the odd sites $|\psi_0^{\mathrm{o}}\rangle = \cc{1} \cc{3}
\ldots \cc{L_\mathrm{S}-1} | 0 \rangle$ are degenerate. They become coupled in
$N$th order in $H_\mathrm{K}$. The degeneracy is therefore lifted even for infinitesimal
$1/U$ and the ground state is given by the symmetric (the effective coupling is
$-(1/U)^{L_\mathrm{S}/2}$) superposition
\begin{equation}
|\psi_0 \rangle = \frac{1}{\sqrt{2}} \left (|\psi_0^{\mathrm{e}}\rangle +
|\psi_0^{\mathrm{o}}\rangle \right ) .
\label{gs}
\end{equation}
In the leads, $H_\mathrm{K}$ yields plane wave eigenstates with the boundary condition
that, due to the Mott-Hubbard gap (of the order of $U$), their wave-function
vanishes (like $1/U$) on the first interacting site. For leads of length $L$ on
either side of the system this gives one-body states with energy $\epsilon_k =
-2 \cos{k}$ and wave-functions $\langle i \ket{k^\mathrm{left}} = \sqrt{2/L}
\sin{k(1-i)}$ on the left hand side ($i<1$) and $\langle i
\ket{k^\mathrm{right}} = \sqrt{2/L} \sin{k(i-L_\mathrm{S})}$ on the right hand side
($i>L_\mathrm{S}$) of the interacting region.

\begin{figure}[tbp]
\resizebox{1.00\columnwidth}{!}{%
\includegraphics{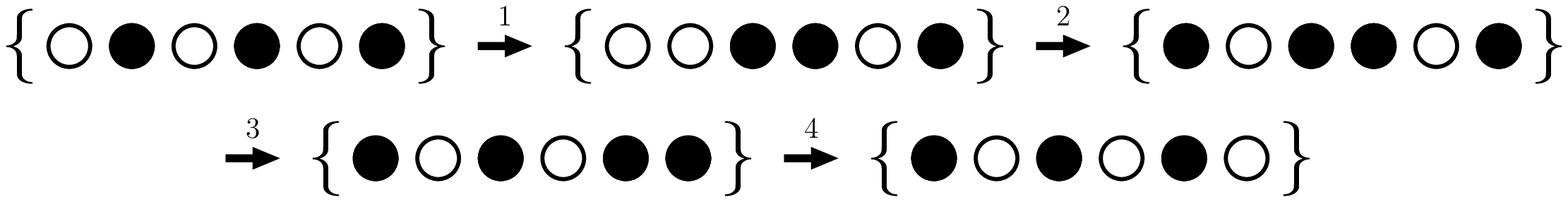}
}
\caption{Example of a hopping sequence of $L_\mathrm{S} / 2 +1=4$ hops
connecting the two components of the ground state (\ref{gs}) of the uncoupled
dot. Full and empty circles represent occupied and empty sites, respectively. In
this example, a particle enters from the left in the second hop, and another one
leaves it to the right in the last hop.} 
\label{sequence} 
\end{figure}

Zero-temperature transport through the interacting segment of the chain requires
that particles are transmitted elastically. Therefore, we need the lowest order
processes in $1/U$ linking the initial state $|i\rangle = \ket{k^\mathrm{left}}
\otimes |\psi_0 \rangle$ to the final state $|f\rangle = |\psi_0 \rangle \otimes
\ket{k^\mathrm{right}}$. These processes are the ones in which a particle is
transmitted while the component $|\psi_0^\mathrm{e}\rangle$ of the ground state
(\ref{gs}) of the interacting region is connected to the component
$|\psi_0^\mathrm{o}\rangle$ by sequences of $N+1$ successive single particle
hoppings. In such a process, each of the $N$ particles inside the interacting
region hops to the next site, and hence the one initially sitting on site
$L_\mathrm{S}$ leaves the region towards the right lead. In addition, a particle
from the left lead enters the dot region and appears on site 1. An example of
such a sequence is sketched in Fig.\ \ref{sequence}.

These dominating processes in the transport through an interacting chain are
similar to the co-tunneling processes through arrays of quantum dots
\cite{aleiner2002}.

The lowest contributions to the effective matrix element $M_{f,i}$ are therefore
of order $N+1$ in $H_\mathrm{K}$. It is given by
\begin{equation}
M_{f,i} = \!\!\!\!\!\!\!\!\! \sum_{S = \left\{ \alpha_1, \alpha_2, \ldots
\alpha_N \right\} } \frac{\langle i | H_\mathrm{K} | \alpha_1 \rangle \langle \alpha_1 |
H_\mathrm{K} | \alpha_2 \rangle \ldots \langle \alpha_N | H_\mathrm{K} | f
\rangle}{( E_i -
E_{\alpha_1})(E_i - E_{\alpha_2})\ldots(E_i - E_{\alpha_N})} ,
\label{Mfi_}
\end{equation}
where $S=\left\{ \alpha_1, \alpha_2, \ldots \alpha_N \right\}$ are sequences of
$N$ intermediate eigenstates of $H_0$ linked by $N+1$ subsequent single particle
hoppings. For all the $(N+1)!$ different permutations of the order of the hops,
the numerator is given by $(-1)^{N+1} \sinsq{k}/L$. The denominator involves $U$
and $V_\mathrm{g}$ and depends on the sequence. For small even $L_\mathrm{S}$
we have explicitly derived the formula
\begin{equation}\label{mmott}
M_{f,i} = - \frac{1}{L}\sinsq{k}\left(\frac{U}{V_\mathrm{g}^2 - U^2/4}\right)^{L_\mathrm{S}/2}
\end{equation} 
and we have confirmed its validity for larger values of $L_\mathrm{S}$ by performing the
summation over the sequences numerically. With (\ref{g}) and the one-dimensional
density of states $\rho = L/(2 \pi \sin{k})$ we get
\begin{equation}
g = \left( \frac{U}{V_\mathrm{g}^2 - U^2/4} \right)^{L_\mathrm{S}}
\label{gg}
\end{equation}
for the dominating contribution to the conductance in the limit of very strong
interaction strength $U$, at $E_\mathrm{F}=0$ ($k_\mathrm{F}=\pi/2$). This
demonstrates that {\it Coulomb blockade eventually occurs for sufficiently
strong interactions}. The suppression of the conductance in the valley due to
the Mott insulating behaviour follows a power law in the interaction strength
$U$ and is exponential in the dot length $L_\mathrm{S}$. The result of Eq.\
(\ref{gg}) is a generalization of the result found for $V_\mathrm{g}=0$ in
\cite{molina2004a}. The small values of the conductance around $V_\mathrm{g}=0$
in Fig.\ \ref{all}c are well described by Eq.\ (\ref{gg}) and the agreement
improves for stronger interaction.

The power-law decay in Eq.\ (\ref{gg}) of the valley conductance with very large
interaction strength $U$ is obtained analytically and confirmed by very accurate
numerics in our case of spinless fermions. For the case of Hubbard chains, an
exponential decay was concluded \cite{oguri2005} from NRG data for relatively
moderate values of the interaction.  However, our analytical approach can be
applied to the Hubbard case as well, leading to a power law dependence of the
conductance on the interaction strength in the limit of large interaction
strength. This power law is a consequence of the lattice model. The exponential
decay mentioned in \cite{oguri2005} might be an apparent dependence in a regime
of intermediate interaction strength.

Within the same model, but in the absence of a gate voltage, it was previously
found \cite{molina2003} that the smoothing of the contacts (i.e. the slow
branching of the interaction going from the leads to the dot) reduces the
interaction-induced suppression of the conductance found for even values of
$L_\mathrm{S}$. In the framework of the present paper, this means that smoothing the
contacts will reduce the depth of the conductance valley at zero gate voltage.
However, an adiabatic branching of the interaction is necessary to recover the
perfect conductance found without interaction. Furthermore, numerical results
\cite{vasseur_thesis} show that smoothing the interaction does not necessarily
suppress the other valleys. 

\subsection{Peak positions and shapes}
\label{peaks}

\begin{figure}[tbp]
\resizebox{0.95\columnwidth}{!}{%
\includegraphics{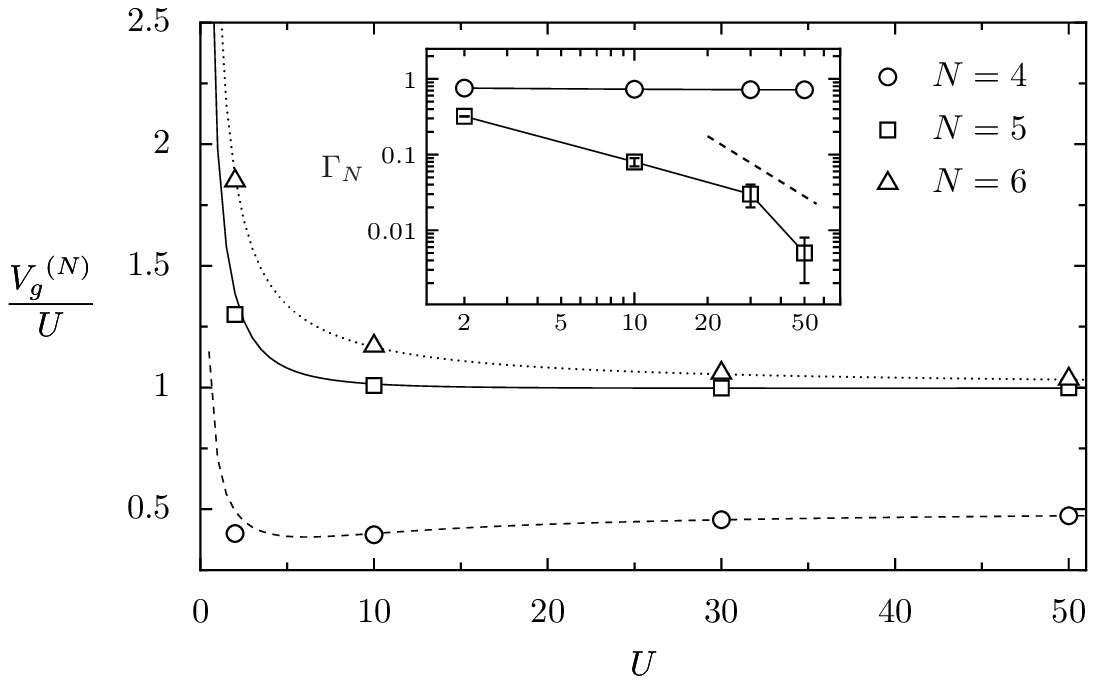}
}
\caption{Positions ${V_\mathrm{g}}^{(N)}/U$ of the peaks from the numerically
obtained conductance (as in Fig.\ \ref{all}) as a function of the
interaction strength $U$ (symbols). The lines give the isolated dot prediction
${\widetilde{V}_\mathrm{g}}^{(N)} / U$ according to Eq.\ (\ref{pred}). The inset
shows the peak width $\Gamma_N$ as a function of $U$ in double logarithmic
scale. The error bars are given by the uncertainty of the fitting. For $N=4$
they are smaller than the symbol size. The $N=6$ data points (not shown) are the
lowest and exhibit the largest error bars. The dashed line represents the
asymptotic power law $1/U^2$.
}
\label{ps} 
\end{figure}

From the data presented on Fig.\ \ref{all}, the positions ${V_\mathrm{g}}^{(N)}$ of the
resonances can be determined as the gate voltage values for which there are
precisely $N-1/2$ particles inside the dot. These peak positions
${V_\mathrm{g}}^{(N)}$, in units of $U$, are shown in Fig.\ \ref{ps} (symbols)
as a function of the interaction strength $U$, together with the positions
\begin{equation}
{\widetilde{V}_\mathrm{g}}^{(N)} (U) = E_0 (N, U) - E_0 (N-1, U) ,
\label{pred}
\end{equation}
that are obtained from the many-body energies $E_0 (N, U)$ of the isolated dot
with $N$ particles (lines). One can see that already for $U=10$, the positions of the
resonances are very close to the isolated dot prediction of Eq.\ (\ref{pred}). This
is an additional confirmation of the fact that the interaction leads to a
decoupling of the dot from the leads.

One of the most relevant features of the numerical results shown in Fig.\
\ref{all} is the sharpening of the conductance peaks with increasing
interaction strength. We shall now provide an understanding for the effect of
strong interactions on the peak width.
  
The width $\Gamma_N$ of a peak is given by the value of the matrix element of
$H_\mathrm{C}$ in the numerator of (\ref{Mfi}). Since the conductance resonance
at ${V_\mathrm{g}}^{(N)}$ is given by the degeneracy of the $N$- and
$N-1$-particle ground states, we restrict our discussion to the contribution of
the transition between those two ground states. Within a qualitative discussion,
we can characterize the parameter dependence of the peak width using the
transition matrix element
\begin{equation}
 - \bra{k_\mathrm{F}^\mathrm{left}} \otimes \bra{0(N-1)} \cc{0} c_1 \ket{0(N)},
\end{equation}
which contains the amplitude 
\begin{equation}
\mathscr{C}_N = \bra{0(N-1)} c_1 \ket{0(N)}.
\end{equation}
The main contribution to $\Gamma_N$ is proportional to
$\left|\mathscr{C}_N\right|^2$. This matrix element depends on the many-body
wave-functions, and is therefore strongly influenced by the interactions. For
general values of $N$, it can be highly non-trivial to determine the ground
state of the isolated dot. There are, however, particular values of $N$ for
which we can study $\mathscr{C}_N$ analytically.

One of these special values in the case of even $L_\mathrm{S}$ is $N=
L_\mathrm{S}/2+1$, which corresponds to the left peak in Fig.\ \ref{all}. In the
limit of strong interactions, the $\ket{0(N-1)}$ state is a superposition of two
charge density wave configurations, as presented in Eq.\ (\ref{gs}). The additional particle in
$\ket{0(N)}$ is a defect (with interaction energy $U$) of the Mott insulator
which can propagate freely, and can be described by an
effective one-particle theory. In the absence of the kinetic energy
$H_\mathrm{K}$, the state $\ket{0(N)}$ is a superposition of the $L_\mathrm{S}/2$
degenerate low-lying states of the form $\cc{1} \cc{3} \ldots \cc{2i-1} \cc{2i}
\ldots \cc{L_\mathrm{S}} \ket{0}$. In the presence of $H_\mathrm{K}$, the
weights of the different components are obtained by solving the equivalent
problem of a free particle on $L_\mathrm{S}/2$ sites. Only the weight of the
first state (namely for $i=1$) enters the expression for
$\mathscr{C}_{L_\mathrm{S}/2+1}$, and one obtains
\begin{equation}
\mathscr{C}_{L_\mathrm{S}/2+1} = \frac{1}{\sqrt{L_\mathrm{S}/2 + 1}} \sin{\frac{\pi}{L_\mathrm{S}/2
    + 1}}.
\end{equation}
In this limit of strong interaction, the lowest order contribution to
$\mathscr{C}_{L_\mathrm{S}/2+1}$ is thus independent of $U$ (the limiting value is reached
already for $U\gtrsim 2$ as can be seen in the inset of Fig.\ \ref{ps}), but
decreases as $L_\mathrm{S}^{-3/2}$ with the length of the system when
$L_\mathrm{S}$ is large, indicating a decrease of the peak width proportional to
$L_\mathrm{S}^{-3}$. 

It is of some interest to compare this behaviour at large $L_\mathrm{S}$ with a
non-interacting evaluation of the matrix element which one can calculate
although at $U=0$ the dot is not decoupled from the leads. In this
non-interacting case, the eigenstates are given by free particles on $L_\mathrm{S}$ sites.
$\mathscr{C}_{L_\mathrm{S}/2+1}$ is then obtained from the value of the ($L_\mathrm{S}/2+1$)th
lowest one-body wave-function on the first site. This leads to a size dependence
of the matrix element as $L_\mathrm{S}^{-1/2}$ only. From this, one may conclude that the
interactions play an important role in narrowing the resonance. 

The other case in which analytical calculations are straightforward is the one
of the last peak when $N=L_\mathrm{S}$. In this case $\ket{0(N)}$ corresponds to the
full dot. The state $\ket{0(N-1)}$ corresponds to one hole in a potential well
with $U/2$ steps on the edges (as sketched in Fig.\ \ref{system}). It is obvious
that when $U$ increases, the hole is less likely to be found on an edge site,
because of the potential step. The ground state of such a one-body system is
solved by the ansatz of an even wave-function with $\psi(1<n<L_\mathrm{S}) = A
\cos{k(n-(L_\mathrm{S}+1)/2)}$ and $\psi(1)=\psi(L_\mathrm{S})=B$. The
normalization allows to get $B$, which yields
\begin{equation}\label{c_ls-1}
\mathscr{C}_{L_\mathrm{S}} = \frac{\cos{\frac{L_\mathrm{S}-1}{2} k}}{\sqrt{
\sum_{n=1}^{L_\mathrm{S}}\cossq{\left(n -\frac{L_\mathrm{S} +1}{2}\right) k}}},
\end{equation}
where $k$ is the smallest positive value that satisfies the condition
\begin{equation}\label{c_ls-1_condition}
\cos{\frac{L_\mathrm{S}+1}{2} k} = \frac{-U}{2} \cos{\frac{L_\mathrm{S}-1}{2} k} .
\end{equation}
In this case, it can be seen that $\mathscr{C}$ depends on $U$ via $k$. In the
limit of strong interaction, (\ref{c_ls-1_condition}) can only be fulfilled if
$\cos{\frac{L_\mathrm{S}-1}{2} k}\sim U^{-1}$. This implies that the numerator of
(\ref{c_ls-1}) decreases proportionally to $U^{-1}$ and thus the peak width
$\Gamma_N$ is expected to decrease as $U^{-2}$. This impressive example of the
reduction of the peak width due to the interaction is illustrated in the inset
of the Fig.\ \ref{ps}.

We end this section with a discussion on the symmetry of the peaks. As can be
seen in Fig.\ \ref{all}c, the conductance peaks are not of symmetric Lorentzian
shape. This is most obvious for the first peak and can be explained by an
asymmetry in the coupling of the many-body states to higher and lower mean
particle numbers. In order to understand this, we consider the representation of
the matrix element $M_{f,i}$ as a single sum over the eigenstates $\ket{\alpha}$
of the isolated dot, as in Eq.\ (\ref{Mfi}).

Outside resonances, the number $N$ of particles inside the dot, and therefore
the initial (\ref{i}) and final (\ref{f}) states, are well-defined, and one can
use Eq.\ (\ref{g}) to evaluate the conductance. Very close to the conductance
peak however, the transmission becomes large and the Fermi golden rule approach
is not valid. To address the issue of the asymmetry of the peak, we consider
values of the gate voltage which are in the vicinity of a resonance ${V_\mathrm{g}}^{(N)}$ (and away
from all the other), but far enough from the peak centre to have a small
conductance. In this way, the conductance in the tails of the peaks can be
treated using (\ref{Mfi}) and (\ref{g}). 

The main contributions to the sum (\ref{Mfi}) depend on whether the gate voltage
is above or below the resonance value ${V_\mathrm{g}}^{(N)}$. This resonance
corresponds to the transition from $N-1$ to $N$ particles in the dot.

For gate voltage values below the resonance, the states $\ket{i}$ and $\ket{f}$
are $N-1$-particle states, and the most important contributions to $M_{f,i}$
come from the sum over the states $\ket{\alpha}$ with $N$ particles inside the
dot. The terms with $N-2$ particle states $\ket{\alpha}$ are suppressed by the
much larger energy denominator in (\ref{Mfi}). For gate voltages slightly above
${V_\mathrm{g}}^{(N)}$, the initial and final states are $N$-particle states and
the main contributions to the sum are due to intermediate $N-1$-particle states.

Since the many-body excitation spectrum of the isolated dot can depend
dramatically on $N$, the sum can give very different results on the two sides of
the peaks, which are therefore expected not to be symmetric.

We illustrate this mechanism for the example of the first peak of 
Fig.\ \ref{all}c, corresponding to the transition between half filling,
$N-1=L_\mathrm{S}/2=3$, and $N=4$. On the left-hand side of this peak, the sum is
dominated by the intermediate $4$-particle eigenstates of the isolated dot,
while the evaluation of the conductance on the right-hand side of the peak is
dominated by the $3$-particle eigenstates of the dot. The $3$-particle
spectrum of the dot is characterized by the Mott-gap which is of the order of
$U$.  The contribution of the excited $3$-particle states is therefore
suppressed by energy denominators of the order of $U$, leading to a smaller
value of $M_{f,i}$ on the right-hand side of the peak than on the left-hand
side, thus explaining the peak asymmetry observed on Fig.\ \ref{all}c.

\section{Conclusion
\label{conclusion}}

We have shown that Coulomb blockade physics can occur when an interacting system
is coupled to leads by perfect contacts. The underlying mechanism is that while
the contacts are perfect on the one-particle level, the interaction can
introduce many-body effects which effectively reduce the coupling between the
interacting system and the leads.

In order to illustrate this peculiar behaviour, we have studied the gate-voltage
dependence of the conductance through a one-dimensional chain in which spinless
fermions are strongly interacting. We have used the embedding method to compute
the conductance $g$ and the mean number $N$ of electrons inside the interacting
chain from numerical DMRG data for the many-body ground state properties of
large rings. This approach takes into account all the electronic correlations.
Despite the absence of potential barriers separating the interacting chain from
the leads, the numerical results in the regime of strong interaction
unambiguously show features which are characteristic of Coulomb blockade physics
in the transport through quantum dots. Peaks appear in the gate-voltage
dependence of $g$, which are accompanied by steps in $N$. These structures
become more and more pronounced, and the conductance becomes more and more
suppressed in the valleys between the peaks, as the interaction strength is
increased.

Our results indicate that, in spite of the perfect contacts, many-body effects
lead to an effective decoupling of the interacting chain from the leads when the
interaction is sufficiently strong. One can then consider the interacting part
of the chain as a quantum dot. In order to interpret these results, we presented
an analytical study of the model. We explained the deepening of the valleys (and
therefore the occurrence of Coulomb blockade itself) from a perturbative
calculation of the conductance in the regime of strong interaction. In addition,
we showed that the positions of the peaks can be deduced from the many-body
eigenenergies of the isolated dot at strong interaction, confirming that the dot
becomes effectively decoupled from the leads. Furthermore, we explained the
narrowing of the width of the peaks with the interaction strength, as well as
the asymmetry of the peak shape, by analyzing the effective matrix elements for
the dominating transitions.

The current tendency towards the development of smaller MOSFETs
\cite{boehm2005,max} is likely to lead to regimes where the theoretical concepts
discussed in this work could be properly applied. It would be interesting to
further develop our model in order to bridge the gap with those experiments.
Addressing the effects of finite temperature, longer range interactions, and
electron spin are envisioned, as well as going beyond the one-dimensional case.
The newly developed method of obtaining non-linear conductances from
time-dependent DMRG \cite{schneiderCM} could be applied to our model, and it
would be interesting to study the suppression of Coulomb blockade by the bias
voltage.

\begin{acknowledgement}
We acknowledge useful discussions with M.\ Hofheinz, G.-L.\ Ingold, R.\ A.\
Molina, J.-L.\ Pichard and M.\ Sanquer. We thank P.\ Schmitteckert for
discussions and for providing his DMRG code. 
\end{acknowledgement}

\bibliography{bibliography,article}

\begin{thebibliography}{10}
\expandafter\ifx\csname natexlab\endcsname\relax\def\natexlab#1{#1}\fi
\expandafter\ifx\csname bibnamefont\endcsname\relax
  \def\bibnamefont#1{#1}\fi
\expandafter\ifx\csname bibfnamefont\endcsname\relax
  \def\bibfnamefont#1{#1}\fi
\expandafter\ifx\csname citenamefont\endcsname\relax
  \def\citenamefont#1{#1}\fi
\expandafter\ifx\csname url\endcsname\relax
  \def\url#1{\texttt{#1}}\fi
\expandafter\ifx\csname urlprefix\endcsname\relax\def\urlprefix{URL }\fi
\providecommand{\bibinfo}[2]{#2}
\providecommand{\eprint}[2][]{\url{#2}}

\bibitem[{1}]{kouwenhoven1997}
\bibinfo{author}{\bibfnamefont{L.~P.} \bibnamefont{Kouwenhoven}},
  \bibinfo{author}{\bibfnamefont{C.~M.} \bibnamefont{Marcus}},
  \bibinfo{author}{\bibfnamefont{P.~L.} \bibnamefont{McEuen}},
  \bibinfo{author}{\bibfnamefont{S.}~\bibnamefont{Tarucha}},
  \bibinfo{author}{\bibfnamefont{R.~M.} \bibnamefont{Westervelt}},
  \bibnamefont{and} \bibinfo{author}{\bibfnamefont{N.~S.}
  \bibnamefont{Wingreen}}, in \emph{\bibinfo{booktitle}{Mesoscopic electron
  transport}}, edited by \bibinfo{editor}{\bibfnamefont{L.~L.}
  \bibnamefont{Sohn}}, \bibinfo{editor}{\bibfnamefont{L.~P.}
  \bibnamefont{Kouwenhoven}}, \bibnamefont{and}
  \bibinfo{editor}{\bibfnamefont{G.}~\bibnamefont{Sch{\"o}n}}
  (\bibinfo{year}{1997}).

\bibitem[{2}]{beenakker1991}
\bibinfo{author}{\bibfnamefont{C.~W.~J.} \bibnamefont{Beenakker}},
  \bibinfo{journal}{Phys. Rev. B} \textbf{\bibinfo{volume}{44}},
  \bibinfo{pages}{1646} (\bibinfo{year}{1991}).

\bibitem[{3}]{alphenaar1992}
\bibinfo{author}{\bibfnamefont{B.~W.} \bibnamefont{Alphenaar}},
  \bibinfo{author}{\bibfnamefont{A.~A.~M.} \bibnamefont{Staring}},
  \bibinfo{author}{\bibfnamefont{H.}~\bibnamefont{van Houten}},
  \bibinfo{author}{\bibfnamefont{M.~A.~A.} \bibnamefont{Mabesoone}},
  \bibinfo{author}{\bibfnamefont{O.~J.~A.} \bibnamefont{Buyk}},
  \bibnamefont{and} \bibinfo{author}{\bibfnamefont{C.~T.} \bibnamefont{Foxon}},
  \bibinfo{journal}{Phys. Rev. B} \textbf{\bibinfo{volume}{46}},
  \bibinfo{pages}{7236} (\bibinfo{year}{1992}).

\bibitem[{4}]{marmorkos1992}
\bibinfo{author}{\bibfnamefont{I.~K.} \bibnamefont{Marmorkos}}
  \bibnamefont{and} \bibinfo{author}{\bibfnamefont{C.~W.~J.}
  \bibnamefont{Beenakker}}, \bibinfo{journal}{Phys. Rev. B}
  \textbf{\bibinfo{volume}{46}}, \bibinfo{pages}{15562} (\bibinfo{year}{1992}).

\bibitem[{5}]{aleiner2002}
\bibinfo{author}{\bibfnamefont{I.~L.} \bibnamefont{Aleiner}},
  \bibinfo{author}{\bibfnamefont{P.~W.} \bibnamefont{Brouwer}},
  \bibnamefont{and} \bibinfo{author}{\bibfnamefont{L.~I.}
  \bibnamefont{Glazman}}, \bibinfo{journal}{Phys. Rep.}
  \textbf{\bibinfo{volume}{358}}, \bibinfo{pages}{309} (\bibinfo{year}{2002}).

\bibitem[{6}]{brouwer2005}
\bibinfo{author}{\bibfnamefont{P.~W.} \bibnamefont{Brouwer}},
  \bibinfo{author}{\bibfnamefont{A.}~\bibnamefont{Lamacraft}},
  \bibnamefont{and}
  \bibinfo{author}{\bibfnamefont{K.}~\bibnamefont{Flensberg}},
  \bibinfo{journal}{Phys. Rev. B} \textbf{\bibinfo{volume}{72}},
  \bibinfo{pages}{075316} (\bibinfo{year}{2005}).

\bibitem[{7}]{nazarov1999}
\bibinfo{author}{\bibfnamefont{Y.~V.} \bibnamefont{Nazarov}},
  \bibinfo{journal}{Phys. Rev. Lett.} \textbf{\bibinfo{volume}{82}},
  \bibinfo{pages}{1245} (\bibinfo{year}{1999}).

\bibitem[{8}]{golubev2004}
\bibinfo{author}{\bibfnamefont{D.~S.} \bibnamefont{Golubev}} \bibnamefont{and}
  \bibinfo{author}{\bibfnamefont{A.~D.} \bibnamefont{Zaikin}},
  \bibinfo{journal}{Phys. Rev. B} \textbf{\bibinfo{volume}{69}},
  \bibinfo{pages}{075318} (\bibinfo{year}{2004}).

\bibitem[{9}]{huibers1998}
\bibinfo{author}{\bibfnamefont{A.~G.} \bibnamefont{Huibers}},
  \bibinfo{author}{\bibfnamefont{S.~R.} \bibnamefont{Patel}},
  \bibinfo{author}{\bibfnamefont{C.~M.} \bibnamefont{Marcus}},
  \bibinfo{author}{\bibfnamefont{P.~W.} \bibnamefont{Brouwer}},
  \bibinfo{author}{\bibfnamefont{C.~I.} \bibnamefont{Duru{\"o}z}},
  \bibnamefont{and} \bibinfo{author}{\bibfnamefont{J.~S.} \bibnamefont{{Harris,
  Jr.}}}, \bibinfo{journal}{Phys. Rev. Lett.} \textbf{\bibinfo{volume}{81}},
  \bibinfo{pages}{1917} (\bibinfo{year}{1998}).

\bibitem[{10}]{boehm2005}
\bibinfo{author}{\bibfnamefont{M.}~\bibnamefont{Boehm}},
  \bibinfo{author}{\bibfnamefont{M.}~\bibnamefont{Hofheinz}},
  \bibinfo{author}{\bibfnamefont{X.}~\bibnamefont{Jehl}},
  \bibinfo{author}{\bibfnamefont{M.}~\bibnamefont{Sanquer}},
  \bibinfo{author}{\bibfnamefont{M.}~\bibnamefont{Vinet}},
  \bibinfo{author}{\bibfnamefont{B.}~\bibnamefont{Previtali}},
  \bibinfo{author}{\bibfnamefont{B.}~\bibnamefont{Fraboulet}},
  \bibinfo{author}{\bibfnamefont{D.}~\bibnamefont{Mariolle}}, \bibnamefont{and}
  \bibinfo{author}{\bibfnamefont{S.}~\bibnamefont{Deleonibus}},
  \bibinfo{journal}{Phys. Rev. B} \textbf{\bibinfo{volume}{71}},
  \bibinfo{pages}{033305} (\bibinfo{year}{2005}).

\bibitem[{11}]{max}
\bibinfo{note}{M. Hofheinz, X. Jehl, M. Sanquer, G. Molas, M. Vinet and S.
  Deleonibus, arXiv:cond-mat/0504325.}

\bibitem[{12}]{nisikawa_cond-mat}
\bibinfo{note}{Y. Nisikawa and A. Oguri, arXiv:cond-mat/0511182}.

\bibitem[{13}]{bohr2006}
\bibinfo{author}{\bibfnamefont{D.}~\bibnamefont{Bohr}},
  \bibinfo{author}{\bibfnamefont{P.}~\bibnamefont{Schmitteckert}},
  \bibnamefont{and}
  \bibinfo{author}{\bibfnamefont{P.}~\bibnamefont{W{\"o}lfle}},
  \bibinfo{journal}{Europhys. Lett.} \textbf{\bibinfo{volume}{73}},
  \bibinfo{pages}{246} (\bibinfo{year}{2006}).

\bibitem[{14}]{schneiderCM}
\bibinfo{note}{G. Schneider and P. Schmitteckert, arXiv:cond-mat/0601389}.

\bibitem[{15}]{blanter1997}
\bibinfo{author}{\bibfnamefont{Y.~M.} \bibnamefont{Blanter}},
  \bibinfo{author}{\bibfnamefont{A.~D.} \bibnamefont{Mirlin}},
  \bibnamefont{and} \bibinfo{author}{\bibfnamefont{B.~A.}
  \bibnamefont{Muzykantskii}}, \bibinfo{journal}{Phys. Rev. Lett.}
  \textbf{\bibinfo{volume}{78}}, \bibinfo{pages}{2450} (\bibinfo{year}{1997}).

\bibitem[{16}]{cohen1999}
\bibinfo{author}{\bibfnamefont{A.}~\bibnamefont{Cohen}},
  \bibinfo{author}{\bibfnamefont{K.}~\bibnamefont{Richter}}, \bibnamefont{and}
  \bibinfo{author}{\bibfnamefont{R.}~\bibnamefont{Berkovits}},
  \bibinfo{journal}{Phys. Rev. B} \textbf{\bibinfo{volume}{60}},
  \bibinfo{pages}{2536} (\bibinfo{year}{1999}).

\bibitem[{17}]{molina2003}
\bibinfo{author}{\bibfnamefont{R.~A.} \bibnamefont{Molina}},
  \bibinfo{author}{\bibfnamefont{D.}~\bibnamefont{Weinmann}},
  \bibinfo{author}{\bibfnamefont{R.~A.} \bibnamefont{Jalabert}},
  \bibinfo{author}{\bibfnamefont{G.-L.} \bibnamefont{Ingold}},
  \bibnamefont{and} \bibinfo{author}{\bibfnamefont{J.-L.}
  \bibnamefont{Pichard}}, \bibinfo{journal}{Phys. Rev. B}
  \textbf{\bibinfo{volume}{67}}, \bibinfo{pages}{235306}
  (\bibinfo{year}{2003}).

\bibitem[{18}]{molina2004a}
\bibinfo{author}{\bibfnamefont{R.~A.} \bibnamefont{Molina}},
  \bibinfo{author}{\bibfnamefont{P.}~\bibnamefont{Schmitteckert}},
  \bibinfo{author}{\bibfnamefont{D.}~\bibnamefont{Weinmann}},
  \bibinfo{author}{\bibfnamefont{R.~A.} \bibnamefont{Jalabert}},
  \bibinfo{author}{\bibfnamefont{G.-L.} \bibnamefont{Ingold}},
  \bibnamefont{and} \bibinfo{author}{\bibfnamefont{J.-L.}
  \bibnamefont{Pichard}}, \bibinfo{journal}{Eur. Phys. J. B}
  \textbf{\bibinfo{volume}{39}}, \bibinfo{pages}{107} (\bibinfo{year}{2004}).

\bibitem[{19}]{future}
\bibinfo{note}{G. Vasseur, D. Weinmann and R. A. Jalabert, unpublished.}

\bibitem[{20}]{molina2005}
\bibinfo{author}{\bibfnamefont{R.~A.} \bibnamefont{Molina}},
  \bibinfo{author}{\bibfnamefont{D.}~\bibnamefont{Weinmann}}, \bibnamefont{and}
  \bibinfo{author}{\bibfnamefont{J.-L.} \bibnamefont{Pichard}},
  \bibinfo{journal}{Eur. Phys. J. B} \textbf{\bibinfo{volume}{48}},
  \bibinfo{pages}{243} (\bibinfo{year}{2005}).

\bibitem[{21}]{rejec2003}
\bibinfo{author}{\bibfnamefont{T.}~\bibnamefont{Rejec}} \bibnamefont{and}
  \bibinfo{author}{\bibfnamefont{A.}~\bibnamefont{Ram{\v s}ak}},
  \bibinfo{journal}{Phys. Rev. B} \textbf{\bibinfo{volume}{68}},
  \bibinfo{pages}{035342} (\bibinfo{year}{2003}).

\bibitem[{22}]{favand1998}
\bibinfo{author}{\bibfnamefont{J.}~\bibnamefont{Favand}} \bibnamefont{and}
  \bibinfo{author}{\bibfnamefont{F.}~\bibnamefont{Mila}},
  \bibinfo{journal}{Eur. Phys. J. B} \textbf{\bibinfo{volume}{2}},
  \bibinfo{pages}{293} (\bibinfo{year}{1998}).

\bibitem[{23}]{sushkov2001}
\bibinfo{author}{\bibfnamefont{O.~P.} \bibnamefont{Sushkov}},
  \bibinfo{journal}{Phys. Rev. B} \textbf{\bibinfo{volume}{64}},
  \bibinfo{pages}{155319} (\bibinfo{year}{2001}).

\bibitem[{24}]{meden2003}
\bibinfo{author}{\bibfnamefont{V.}~\bibnamefont{Meden}} \bibnamefont{and}
  \bibinfo{author}{\bibfnamefont{U.}~\bibnamefont{Schollw{\"o}ck}},
  \bibinfo{journal}{Phys. Rev. B} \textbf{\bibinfo{volume}{67}},
  \bibinfo{pages}{193303} (\bibinfo{year}{2003}).

\bibitem[{25}]{DMRG_book}
\bibinfo{note}{{\textit{Density-Matrix Renormalization {--} A New Numerical
  Method in Physics}}, ed. by I. Peschel, X. Wang, M. Kaulke and K. Hallberg,
  Springer (Berlin/Heidelberg, 1999).}

\bibitem[{26}]{schmitteckert_thesis}
\bibinfo{note}{P. Schmitteckert, Ph.D. thesis (Univ. Augsburg, 1996).}

\bibitem[{27}]{friedel1}
\bibinfo{author}{\bibfnamefont{J.}~\bibnamefont{Friedel}},
  \bibinfo{journal}{Phil. Mag.} \textbf{\bibinfo{volume}{43}},
  \bibinfo{pages}{153} (\bibinfo{year}{1952}).

\bibitem[{28}]{friedel2}
\bibinfo{note}{J. S. Langer and V. Ambegaokar, Phys. Rev. {\textbf{121}}, 1090
  (1961); D. C. Langreth, Phys. Rev. {\textbf{150}}, 516 (1966).}

\bibitem[{29}]{molina2004}
\bibinfo{author}{\bibfnamefont{R.~A.} \bibnamefont{Molina}},
  \bibinfo{author}{\bibfnamefont{D.}~\bibnamefont{Weinmann}}, \bibnamefont{and}
  \bibinfo{author}{\bibfnamefont{J.-L.} \bibnamefont{Pichard}},
  \bibinfo{journal}{Europhys. Lett.} \textbf{\bibinfo{volume}{67}},
  \bibinfo{pages}{96} (\bibinfo{year}{2004}).

\bibitem[{30}]{oguri1999}
\bibinfo{author}{\bibfnamefont{A.}~\bibnamefont{Oguri}},
  \bibinfo{journal}{Phys. Rev. B} \textbf{\bibinfo{volume}{59}},
  \bibinfo{pages}{12240} (\bibinfo{year}{1999}).

\bibitem[{31}]{smit2003}
\bibinfo{author}{\bibfnamefont{R.~H.~M.} \bibnamefont{Smit}},
  \bibinfo{author}{\bibfnamefont{C.}~\bibnamefont{Untiedt}},
  \bibinfo{author}{\bibfnamefont{G.}~\bibnamefont{{Rubio-Bollinger}}},
  \bibinfo{author}{\bibfnamefont{R.~C.} \bibnamefont{Segers}},
  \bibnamefont{and} \bibinfo{author}{\bibfnamefont{J.~M.} \bibnamefont{{van
  Ruitenbeek}}}, \bibinfo{journal}{Phys. Rev. Lett.}
  \textbf{\bibinfo{volume}{91}}, \bibinfo{pages}{076805}
  (\bibinfo{year}{2003}).

\bibitem[{32}]{selva2000}
\bibinfo{author}{\bibfnamefont{F.}~\bibnamefont{Selva}} \bibnamefont{and}
  \bibinfo{author}{\bibfnamefont{D.}~\bibnamefont{Weinmann}},
  \bibinfo{journal}{Eur. Phys. J. B} \textbf{\bibinfo{volume}{18}},
  \bibinfo{pages}{137} (\bibinfo{year}{2000}).

\bibitem[{33}]{oguri2005}
\bibinfo{author}{\bibfnamefont{A.}~\bibnamefont{Oguri}} \bibnamefont{and}
  \bibinfo{author}{\bibfnamefont{A.~C.} \bibnamefont{Hewson}},
  \bibinfo{journal}{J. Phys. Soc. Jpn.} \textbf{\bibinfo{volume}{74}},
  \bibinfo{pages}{988} (\bibinfo{year}{2005}).

\bibitem[{34}]{vasseur_thesis}
\bibinfo{note}{G. Vasseur, Ph.D. thesis, in preparation (ULP Strasbourg,
  2006).}

\end{thebibliography}
\end{document}